\documentstyle[aps,epsf,prl,epsfig,twocolumn]{revtex}
\def\be{\begin{equation}}
\def\ee{\end{equation}}
\def\bea{\begin{eqnarray}}          
\def\eea{\end{eqnarray}}
\def\bi{\begin{itemize}}
\def\ei{\end{itemize}}

\begin{document}

\title{ Dynamics of a quantum phase transition in the random Ising model }

\author{ Jacek Dziarmaga }

\address{ Institute of Physics and Centre for Complex Systems,
          Jagiellonian University,
          Reymonta 4, 30-059 Krak\'ow, Poland  }

\date{ April 5, 2006 }

\maketitle

\begin{abstract}
A quantum phase transition from paramagnetic to ferromagnetic phase 
is driven by a time-dependent external magnetic field. For any rate of 
the transition the evolution is non-adiabatic and finite density of 
defects is excited in the ferromagnetic state. The density of excitations
has only logarithmic dependence on the transition rate. This is much
weaker than any usual power law scaling predicted for pure systems 
by the Kibble-Zurek mechanism.
\end{abstract}

PACS numbers: 03.65.-w, 73.43.Nq, 03.75.Lm, 32.80.Bx, 05.70.Fh 

According to the book \cite{book} our understanding of quantum phase transitions is 
based on two prototypical models: the quantum Ising chain and the Bose-Hubbard model.  
It is well established in the exactly solvable Ising model \cite{Dziarmaga2005,KZIsing,LZ} 
and there are indications in the Bose-Hubbard model \cite{KZBH} that essential properties 
of a dynamical quantum phase transition are captured by the Kibble-Zurek (KZ) mechanism
of defect formation \cite{KZM}. This theory was originally developed for thermodynamic 
transitions and tested both by numerical simulations \cite{KZnum} and by experiments in 
finite-temperature condensed matter systems \cite{tests}. It is based on the observation 
that, because of the critical slowing down, a system driven across its critical point 
must go out of thermal equilibrium no matter how slow is the transition rate. Using only
combination of general casuality and universality arguments it shows that in a 
transition from a disordered to ordered phase the system ends in a non-equilibrium 
state with finite ordered domains of average size
\be
\hat\xi~\simeq~\tau_Q^{\frac{\nu}{z\nu+1}}~.
\label{xiKZ}
\ee
Here $z$ and $\nu$ are critical exponents and $\tau_Q$ is time of the transition 
(or quench). No matter how slow is the transition the system does not have enough time 
to order throughout its whole volume and average size of ordered domains is limited to a 
finite $\hat\xi$ which is a power of the transition time $\tau_Q$. 

The short list of the two prototypical models must be supplemented by a prototypical 
disordered system i.e. the random Ising chain defined by the Hamiltonian 
\cite{random,DFisher}
\be
H~=~-\sum_{n=1}^N \left( h~\sigma^x_n + J_n~\sigma^z_n\sigma^z_{n+1} \right)~.
\label{Hsigma}
\ee 
with periodic boundary conditions 
\be
\vec\sigma_{N+1}~=~\vec\sigma_1~.
\ee 
Here $J_n$'s are random ferromagnetic couplings, $J_n>0$, and $h$ is external magnetic 
field. This model has a quantum critical point at 
$h_c=\exp\left(\overline{\ln J_n}\right)$ separating ferromagnetic ($h<h_c$) from  
paramagnetic ($h>h_c$) phase. When $h\to\infty$ the ground state becomes fully polarized 
along the $x$-axis,
\be
\left|
\rightarrow\rightarrow\rightarrow\rightarrow\rightarrow\rightarrow\dots
\right\rangle~,
\ee
but when $h=0$ there are two degenerate ferromagnetic ground states
\be
\left|
\uparrow\uparrow\uparrow\uparrow\uparrow\uparrow\dots
\right\rangle~,~~~~~
\left|
\downarrow\downarrow\downarrow\downarrow\downarrow\downarrow\dots
\right\rangle~.~
\ee 
The critical point at $h_c$ is surrounded by the Griffith regime (or phase) of 
infinite linear susceptibility. 

The random Ising model is important because real condensed matter spin systems are random. 
Their randomness is not a mere small perturbation on top of pure models because the disorder 
is changing their universality class. For example, the randomness of $J_n$ in the model 
(\ref{Hsigma}) is changing its universality class with respect to the pure Ising chain 
with a constant $J_n=J$. No matter how weak is the randomness of $J_n$ renormalization 
group transformations drive the model towards an infinite disorder fixed point 
\cite{DFisher}. As a result, the random Ising chain has $\nu=2$ instead of $\nu=1$ and, 
more importantly, $z\to\infty$ instead of $z=1$. The diverging dynamical exponent marks 
a qualitative difference between the pure and the disordered model. As the dynamical 
exponent is relevant to dynamical phase transitions, its singularity is suggesting that 
an outcome of the transition in a disordered system can be qualitatively different from 
that in its pure counterpart. This expectation has never been tested in the theory
of dynamical phase transitions.

In the model (\ref{Hsigma}) a dynamical paramagnet-ferromagnet transition is not 
adiabatic and, as a result, the state after transition is a superposition over excited 
states like
\be
\left|
\uparrow\uparrow\uparrow\uparrow\uparrow\uparrow
\downarrow\downarrow\downarrow\downarrow\downarrow\downarrow\downarrow
\uparrow\uparrow\uparrow\uparrow\uparrow
\downarrow\downarrow\downarrow\downarrow\downarrow\dots
\right\rangle
\label{mess}
\ee
with finite ferromagnetic domains separated by domain walls or kinks. The diverging 
dynamical exponent $z$ implies that, in first approximation, the size of the ordered 
domains $\hat\xi$ in Eq.(\ref{xiKZ}) does not depend on the quench time. No matter 
how slow is the transition the average density of kinks $d\sim\hat\xi^{-1}$ remains 
the same.

In second approximation we can follow general principles of the KZ mechanism \cite{KZM} 
to derive $\hat\xi$ when $z\to\infty$. These ideas were originally proposed for classical 
transitions \cite{KZM}, but they were generalized recently to quantum phase transitions 
\cite{KZBH,Bodzio,Dziarmaga2005,KZIsing}. The basic idea is that when a system is driven
through a quantum phase transition, then the evolution of its state is adiabatic when the 
system is away from the critical point, but it must be non-adiabatic in a neighborhood of 
the critical point where the gap $\Delta$ between the ground state and the first excited 
state tends to zero. It is convenient to measure the distance from the critical point by
a dimensionless parameter 
$
\epsilon~=~\frac{h_c-h}{h_c}. 
$
The evolution is non-adiabatic between $-\hat\epsilon$ and $+\hat\epsilon$, where
$\pm\hat\epsilon$ are the two points when transition rate equals the gap $\Delta$. 
On one hand, assuming that $\epsilon(t)$ can be linearized near $\epsilon=0$,
\be
\epsilon(t)~\simeq~\frac{t}{\tau_Q}~,
\ee  
we can estimate the rate as 
$\left|\frac{\dot\epsilon}{\epsilon}\right|\approx\frac{1}{\tau_Q|\epsilon|}$.
On the other hand, we know that for $|\epsilon|\to0$ the gap behaves like
\be
\Delta~\simeq~|\epsilon|^{z\nu}~\simeq~|\epsilon|^{1/|\epsilon|}~,
\label{gap}
\ee
see Ref.\cite{DFisher}. The rate and the gap equal at $\hat\epsilon$ when
\be
\frac{\alpha}{\tau_Q\hat\epsilon}~=~\hat\epsilon^{1/\hat\epsilon}~.
\label{KZadiab}
\ee
Here $\alpha={\cal O}(1)$ is a single fitting parameter. Quick analysis of this
equation shows that $\hat\epsilon\to 0$ when $\tau_Q\to\infty$: for very slow 
transitions the evolution is non-adiabatic only in a very close neighborhood 
of the critical point, where we can use Eq.(\ref{gap}) self-consistently.

In the adiabatic-impulse approximation the state of the system follows its instantaneous
ground state before $-\hat\epsilon$, then its evolution becomes impulse between 
$-\hat\epsilon$ 
and $+\hat\epsilon~$ when the state does not change because reaction of the system is too
slow as compared to the transition rate, and finally the evolution becomes adiabatic again 
after $+\hat\epsilon$. Accuracy of the adiabatic-impulse approximation was tested in the 
quantum context in Refs.\cite{LZ,Bodzio}. The state of the system at $-\hat\epsilon$ is 
the ground state with a finite correlation length 
\be
\hat\xi~=~\frac{1}{\hat\epsilon^\nu}~.
\label{hatxibis}
\ee
This state does not change until $+\hat\epsilon$ when the evolution becomes adiabatic again. 
In this way the correlation length $\hat\xi$ becomes imprinted on the state of the system 
after the transition. This length becomes the average size of ordered domains in the final 
ferromagnetic state (\ref{mess}). For example, in the random Ising model ($\nu=2$) approximate
solution of Eq.(\ref{KZadiab}) in the limit of $\tau_Q\to\infty$ together with 
Eq.(\ref{hatxibis}) gives 
\be
\hat\xi~\sim~\ln^2\tau_Q~
\label{xilog}
\ee 
when $\ln\tau_Q\gg 1$. Again, as in our first approximation, the dependence on 
$\tau_Q$ is very weak as compared to any power law scaling.

As mentioned before, validity of Eq.(\ref{gap}) is limited only to close vicinity of the
critical point. Further away from this point we can expect spectral properties of the model 
to be more like in the pure Ising model. This must be true at least for weak disorder, when 
the random $J_n$'s have a narrow distribution around a finite $J$. As a result, for 
moderately slow transitions, when $\hat\epsilon$ is large enough, we can expect
\be
\hat\xi~\sim~\tau_Q^{1/2}~,
\label{hatxipure}
\ee  
like in the pure Ising model with $\nu=z=1$ \cite{Dziarmaga2005,KZIsing}, see
Eq.(\ref{xiKZ}). 
  
Before proceeding with numerical simulations of a dynamical transition we need 
better understanding of spectral properties of the random Ising chain. To this end 
I assume for convenience that $N$ is even and make the Jordan-Wigner transformation,
\bea
&&
\sigma^x_n~=~1-2 c^\dagger_n  c_n~, \\
&&
\sigma^z_n~=~
-\left( c_n+ c_n^\dagger\right)
 \prod_{m<n}(1-2 c^\dagger_m c_m)~,
\label{JW}
\eea
introducing fermionic operators $c_n$, which satisfy anticommutation relations 
$\left\{c_m,c_n^\dagger\right\}=\delta_{mn}$ and 
$\left\{ c_m, c_n \right\}=\left\{c_m^\dagger,c_n^\dagger \right\}=0$.
The Hamiltonian (\ref{Hsigma}) becomes 
\be
 H~=~P^+~H^+~P^+~+~P^-~H^-~P^-~,
\label{Hc}
\ee
where
\be
P^{\pm}=
\frac12\left[1\pm\prod_{n=1}^N\sigma^x_n\right]=
\frac12\left[1~\pm~\prod_{n=1}^N\left(1-2c_n^\dagger c_n\right)\right]
\label{Ppm}
\ee
are projectors on subspaces with even ($+$) and odd ($-$) numbers of 
$c$-quasiparticles and  
\bea
H^{\pm}&=&
\sum_{n=1}^N
\left( 
h c_n^\dagger  c_n - J_n c_n^\dagger  c_{n+1} - 
J_n c_{n+1}  c_n - \frac{h}{2} 
\right)~+  
\nonumber\\
&&
~{\rm h.c.}~
\label{Hpm}
\eea
are corresponding reduced Hamiltonians. The $c_n$'s in $H^-$ satisfy
periodic boundary conditions $c_{N+1}=c_1$, but the $c_n$'s in $H^+$
must obey $c_{N+1}=-c_1$, what I call ``antiperiodic'' boundary conditions. 

The parity of the number of $c$-quasiparticles is a good quantum number and the ground
state has even parity for any value of $h$. Assuming that the quench begins in the ground 
state we can confine to the subspace of even parity. In this subspace the
quadratic $H^+$ is diagonalized by a Bogoliubov transformation
\be
c_n~=~
\sum_{m=1}^{N} 
\left(
u_{nm}\gamma_m + v^*_{nm} \gamma_m^\dagger
\right)
\label{uv}
\ee 
The index $m$ numbers Bogoliubov modes which are eigenmodes with positive $\omega$ 
of the stationary Bogoliubov-de Gennes equations
\bea
&&
\omega u_n =  \nonumber\\
&&
2 h u_n - 
j_n     \left( u_{n+1} + v_{n+1} \right) +
j_{n-1} \left( v_{n-1} - u_{n-1} \right)~,
\nonumber\\
&&
\omega v_n =  \nonumber\\
&&
- 2 h u_n + 
j_n     \left( v_{n+1} + u_{n+1} \right) +
j_{n-1} \left( v_{n-1} - u_{n-1} \right)~.
\label{BdG}
\eea
Here I suppressed the mode number $m$ and defined $j_n=J_n$ for
$n<N$ and $j_N=-J_N$ to take into account the antiperiodic boundary
conditions. The eigenstates $(u_{nm},v_{nm})$ with positive energy $\omega_m>0$, 
normalized so that $\sum_n\left(|u_{nm}|^2+|v_{nm}|^2\right)=1$, define 
quasiparticle operators $\gamma_m=u_{nm}^*c_n+v_{nm}c_n^\dagger$. Each
positive energy eigenstate has a partner $(u^-_{nm},v^-_{nm})=(v_{nm},u_{nm})$ 
with negative energy $-\omega_m$ which defines a quasiparticle operator 
$\gamma_m^-=(u^-_{nm})^*c_n+v^-_{nm}c_n^\dagger=\gamma_m^\dagger$. 
After the transformation (\ref{uv}) the Hamiltonian 
$H^+=\frac12\sum_{m=1}^N \epsilon_m 
     \left(\gamma_m^\dagger\gamma_m-\gamma_m^{-\dagger}\gamma_m^-\right)$
equivalent to
\be
H^+=
\sum_{m=1}^N \epsilon_m~
\left(
\gamma_m^\dagger \gamma_m-\frac12
\right)~
\label{Hgamma}
\ee
which is a simple-looking sum of quasiparticles. However, thanks to the projection 
$P^+~H^+~P^+$ in Eq.(\ref{Hc}) only states with even numbers of quasiparticles 
belong to the spectrum of $H$.

\begin{figure}[t]
\includegraphics[width=0.95\columnwidth,clip=true]{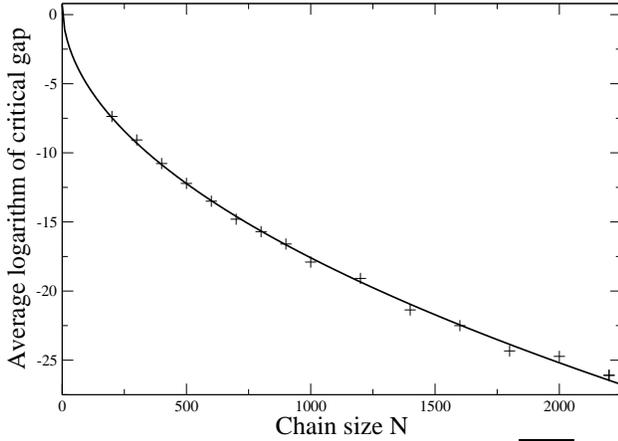}
\caption{ Average logarithm of the critical gap $\overline{\ln\Delta_c}$ as a 
function of $N$. Each circle is an average over $500$ realizations of $J_n$ and the 
solid line is the best fit $\overline{\ln\Delta_c}~=~0.74-0.58\sqrt{N}$. }
\label{fig_gap}
\end{figure}

Infinite Ising chain would have zero energy gap at the critical point but in 
numerical simulations we must use a finite chain with a finite gap $\Delta_c$ 
at $h=h_c$. As the finite gap can alter the adiabaticity condition (\ref{KZadiab}) 
for slow quenches, it is important to know how the gap depends on the system size 
$N$. For example, in the pure Ising chain $\Delta_c$ decays only like $1/N$ and 
relatively large $N$ would be required for accurate numerical simulations. In the random 
Ising chain the ``critical gap'', which is a sum of energies of the two lowest energy
quasiparticles $\Delta_c=\omega_1+\omega_2$ at $h=h_c$, can be found by diagonalization
of Eqs.(\ref{BdG}). The best fit to an average over different realizations of $J_n$ is
\be
\overline{\ln\Delta_c}~=~0.74-0.58\sqrt{N}~, 
\ee
see Figure \ref{fig_gap}. For a given $\tau_Q$ (logarithm of) the critical gap 
is much less than (logarithm of) the transition rate 
$\simeq\tau_Q^{-1}$ provided that $N\gg\ln^2(\tau_Q)$. This condition can be 
easily met on a relatively small lattice. It is not quite surprising that the 
condition is equivalent to $\hat\xi\ll N$, compare Eq.(\ref{xilog}).

Finally, after these preparations, I can proceed with numerical simulations of a dynamical 
quantum phase transition. In my simulations I assumed that the independent random $J_n$'s 
have a uniform distribution in the range $(0,2)$ and the critical field is 
$h_c=\exp\overline{\ln J_n}=\frac{2}{e}$. For the sake of simplicity the transition is 
driven by a linear quench 
\be
h(t)~=~-\frac{t}{\tau_Q}~h_c
\label{quench}
\ee
with $t\in(-\infty,0)$. The system is initially prepared in the ground state at a large 
initial value of the magnetic field $h_i$ i.e. in the Bogoliubov vacuum state for the 
quasiparticles at $h_i$. As the magnetic field is being turned off to zero, the state of 
the system $|\psi(t)\rangle$ is getting excited from its instantaneous ground state. 
However, in a similar way as in Ref.\cite{Dziarmaga2005}, we can follow the time-dependent 
Bogoliubov method and assume that the excited state $|\psi(t)\rangle$ is a Bogoliubov 
vacuum for a set of time-dependent quasiparticle annihilation operators 
\be
\tilde\gamma_m(t)=u_{nm}^*(t)c_n+v_{nm}(t)c_n^\dagger~.
\label{tildegamma}
\ee 
This Ansatz is a solution of Schr\"odinger equation when the Bogoliubov modes $u_{nm}(t)$ 
and $v_{nm}(t)$ solve time-dependent Bogoliubov-de Gennes equations
\bea
&&
i\frac{du_n}{dt} = \nonumber\\
&&
2 h(t) u_n - 
j_n     \left( u_{n+1} + v_{n+1} \right) +
j_{n-1} \left( v_{n-1} - u_{n-1} \right)~,
\nonumber\\
&&
i\frac{dv_n}{dt} = \nonumber\\
&&
- 2 h(t) u_n + 
j_n     \left( v_{n+1} + u_{n+1} \right) +
j_{n-1} \left( v_{n-1} - u_{n-1} \right)~.
\label{tBdG}
\eea
with the initial condition that at $h_i$ each mode is a positive frequency eigenmode 
of the stationary BdG equations (\ref{BdG}). The evolution stops at $h=0$ when the final 
average density of kinks is
\be
d~=~
\frac{1}{N}\sum_n
\left\langle\psi(0)\right|
\frac12\left(1-\sigma^z_n\sigma^z_{n+1}\right)
\left|\psi(0)\right\rangle~.
\ee 
Using the Jordan-Wigner transformation (\ref{JW}) followed by the Bogoliubov 
transformation (\ref{tildegamma}), together with the assumption that
$\tilde\gamma(0)|\psi(0)\rangle=0$, the density of kinks can be transformed into
\be
d=
\frac12-
\frac{1}{N}
\sum_{mn}
\overline{{\rm Re}\left[u_{n+1,m}(0)+v_{n+1,m}(0)\right]v^*_{nm}(0)}~.
\label{d2}
\ee

\begin{figure}[t]
\includegraphics[width=0.95\columnwidth,clip=true]{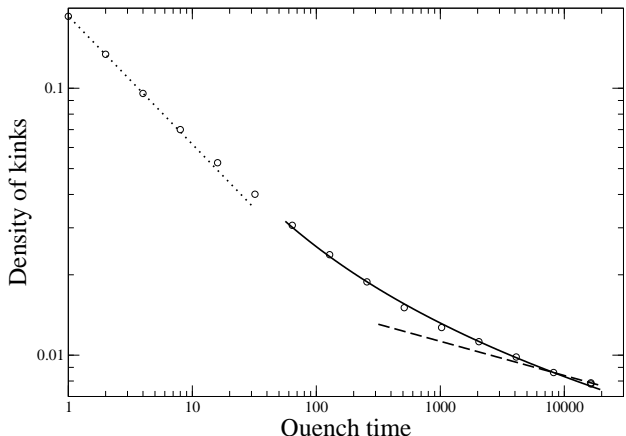}
\caption{ Density of kinks $d$ as a function of transition time $\tau_Q$ on a lattice
of $N=512$ sites. Circles are final kink densities averaged over 4 realizations of $J_n$. 
Error bars set the size of the circles. The solid line is the best 
fit $d(\tau_Q)=\frac{0.15}{\hat{\epsilon}^2(3.4,\tau_Q)}$ where $\hat\epsilon(\alpha,\tau_Q)$ 
is a solution of Eq.(\ref{KZadiab}). The dotted line is the best power 
law fit $d\sim\tau_Q^{-w}$ to the left-most 3 data points and the dashed line is the best 
power law fit to the right-most 2 data points. The exponents are $w=0.48$ and $w=0.13$ 
respectively. The exponent $w=0.48$ for the fastest transitions is consistent with the 
$\frac12$ in the pure Ising model. 
}
\label{fig_kinks}
\end{figure}

The time-dependent BdG equations (\ref{tBdG}) were solved for a range of almost five
decades of $\tau_Q$ with the quench (\ref{quench}) starting at $h_i=10~h_c$. 
Final densities of kinks were collected in Figure \ref{fig_kinks}. 
To test the equation (\ref{KZadiab}), valid for slow transitions, the data for the
slowest quenches were fitted by $d(\tau_Q)=\frac{\beta}{\hat\epsilon^\nu(\alpha,\tau_Q)}$ 
with $\nu=2$. Here $\hat\epsilon(\alpha,\tau_Q)$ is a solution of Eq.(\ref{KZadiab}) 
and $\alpha,\beta$ are two fitting parameters. Both $\alpha=3.02$ and $\beta=0.15$ turned out 
to be ${\cal O}(1)$ and the data are in reasonably good agreement with the best fit.
Although Eq.(\ref{KZadiab}) is confirmed by numerical simulations, even the slowest
quenches with $\tau_Q=16384$, obtained with substantial numerical effort, are not
slow enough to clearly show the asymptotic behavior in Eq.(\ref{xilog}) for 
$\tau_Q\to\infty$. This is not unexpected because this asymptote is achieved when not 
only $\tau_Q\gg 1$ but also $\ln\tau_Q\gg 1$. 

The same figure shows two power-law fits $d\sim\tau_Q^{-w}$. Relatively fast left-most
quenches were fitted with the exponent $w=0.48$ which is close to the $1/2$ in the pure 
Ising model. These quenches, whose $\hat\epsilon$ is relatively large, cannot feel the 
randomness of $J_n$'s and, as predicted, their final defect density scales like 
in the pure Ising model \cite{Dziarmaga2005,KZIsing,LZ}. For comparison, the two right-most 
data points were also fitted by a power law $d\sim\tau_Q^{-w}$, but this time the 
exponent is a mere $w=0.13$. The exponent $w$ decays with increasing $\tau_Q$ 
as expected from the quasi-logarithmic solution $\hat\epsilon(\alpha,\tau_Q)$ of 
Eq.(\ref{KZadiab}).

{\bf Conclusion.---} This paper is the first test of the Kibble-Zurek mechanism 
in a disordered system. The test is passed nicely but with an anomalous result 
that density of defects after a transition depends only logarithmically on 
the transition rate. Crudely speaking, the density is more or less the same no matter 
how slow is the transition. This result is in sharp contrast to the standard power 
law scaling predicted for pure systems. This strongly non-adiabatic behavior can be 
attributed to the anomalously small energy gap near the critical point which is 
characteristic for a random system with an infinite disorder fixed point.

{\bf Acknowledgements. ---} I would like to thank Wojciech Zurek for encouragement,
and Bogdan Damski for comments on the manuscript. This work was supported in part by 
ESF COSLAB programme, US Department of Energy, and Polish Government scientific funds 
(2005-2008) as a research project.

\end{document}